\documentclass[12pt]{article}
\usepackage{epsfig}
\usepackage{amsmath}

\date{\today}

\begin{document}
\title{Chaotic motion in multi-black hole spacetimes 
\\
and  holographic screens}
 \author{{\large {\bf  William Hanan}}  and
{\large {\bf Eugen Radu}}   \\ \\
 {\small Department of
Mathematical Physics,  National University of Ireland, Maynooth, Ireland}\\  
 }
\maketitle 
 
\begin{abstract}
We investigate the geodesic motion in $D-$dimensional Majumdar-Papapetrou multi-black hole
spacetimes and find that the qualitative features of the $D=4$ case  
are shared by the higher dimensional configurations.
The motion of timelike and null particles is chaotic, the phase space being divided into
basins of attraction which are separated by a fractal boundary, with a fractal dimension $d_B$.
The mapping of the geodesic trajectories on a screen
placed in the asymptotic region is also investigated.
We find that the fractal properties of the phase space induces a fractal structure
on the holographic screen, with a fractal dimension $d_B-1$.
\end{abstract}

{\bf Introduction--}
The notion of chaos is one of the most important ideas to explain
various nonlinear phenomena in nature.
An interesting case in which deterministic chaos may occur is given by the
geodesic motion in general relativity.
The nonlinearity of Einstein's equations may give rise to chaos
in systems whose Newtonian analogue is nonchaotic.
Relativistic systems in which chaos is known to appear include charged particles
in a magnetic field interacting with gravitational waves \cite{VP}, particles near a
black hole in a Melvin magnetic universe \cite{KV}, or in
a perturbed Schwarzschild spacetime \cite{Bombelli:1991eg},
\cite{Aguirregabiria:1996vq} 
(see \cite{Sota:1995ms} for a general discussion and other examples).

The geodesic motion in a Majumdar-Papapetrou (MP)
multi-black hole spacetime 
\cite{Majumdar:1947eu,Papaetrou:1947ib} 
provides another particularly interesting situation.
The MP spacetime is a solution of the Einstein-Maxwell system 
in $D-$dimensions, describing 
a collection of $N$ extremal black holes 
located at random in a $D-1$ dimensional hypersurface. Each black hole
has an electric charge equal to its mass, the gravitational and electrostatic
forces canceling pairwise.
As found by several authors \cite{Contopoulos1},
\cite{Dettmann:1994dj}, \cite{Dettmann:1995ex}, the motion of particles for $N=2,~3$
 MP black holes in four dimensions is chaotic, the phase space being divided into
basins of attraction which are separated by a fractal boundary.

In recent years there 
has been a great deal of attention devoted towards research in  
black hole physics in higher dimensions, these 
exhibiting a number of new properties as compared to the $D=4$ case.

The main purpose of this paper is to examine how the chaotic geodesic
motion in a MP multi-black hole background  
depends on the spacetime dimensionality.

In this context, we consider the mapping of the geodesic trajectories
onto a planar screen placed at
a large distance from the black holes.
We find that the fractal properties of the phase space induces a fractal structure
on the holographic screen.

{\bf Chaotic motion in MP spacetime--}
The  MP solution in a $D-$dimensional spacetime has a line element
\begin{eqnarray}
ds^2=-U^{-2}dt^2+U^{\frac{2}{D-3}}\left((dx^{1})^{2}+...+(dx^{D-1})^{2}\right),
\end{eqnarray}
where $U$ is a function of $x^\mu$ 
satisfying the Laplace equations in a $D-1$ dimensional Euclidean space
\begin{eqnarray}
U=1+\sum_{i=1}^{N}\frac{M_{i}}{|r-r_{i}|^{D-3}}~,
\end{eqnarray}
and describes a system of extreme black holes in Einstein-Maxwell theory
with equal charges and masses $M_i>0$.
Here $|r-r_{i}|$ denotes the Euclidean distance between the field point $r$
and the fixed location $r_{i}$ in a Euclidean space:
 $|r-r_{i}|=(\sum_{\mu=1}^{D-1}(x^{\mu}-x_{i}^{\mu})^{2})^{1/2}$. 
The
black hole locations, $r_{i}$, are arbitrary,
the event horizon area of a black hole being
$ M_iA_{D-2}$, where $A_{D-2}$ is the
area of a unit sphere in $D-2$ dimensions.
The Maxwell one-form is
given by ${\cal A}=U \sqrt{(D-2)/(2(D-3))}dt$. 
Note that this solution is static,
with $\partial/\partial t$ a Killing vector, but in general it has no other symmetries.
 
Restricting ourselves to the case of an uncharged test particle 
and considering the contravariant
components  $(\gamma,~{\bf u})$ of the velocity 
 in an orthonormal basis, the geodesic equations of motion are given by
\begin{eqnarray}
\label{geodesics}
\dot{\bf x}&=&U^{-1/(D-3)}{\bf u},~~~\dot{t}=U\gamma,~~~\gamma=\sqrt{\epsilon+u^{2}},
\\
\dot{\bf u}&=&U^{-\frac{D-2}{D-3}}[(\gamma^{2}+\frac{1}{D-3}u^{2})\nabla U
-\frac{1}{D-3}{\bf uu}\cdot\nabla U],
\end{eqnarray}
where  $\epsilon=0,1$ for null and timelike particles respectively;
an overdot indicates the derivative with respect to  $\tau$,
which is an affine parameter along the geodesics ($\tau$ is the  proper time for $\epsilon=1$).
These equations are supplemented with suitable boundary conditions, 
${\bf x}(\tau_0)={\bf x}^{(0)}$, ${\bf u}(\tau_0)={\bf u}^{(0)}$, and possess
a first integral associated with the Killing vector $\partial/\partial t$.

The geodesic equations  have been integrated by using a standard differential
equation solver for spacetime dimensions between four and eight. 
To simplify the general picture, 
the black holes were taken to have  equal masses $M_i=M$,
and were placed equidistantly along the $x^2$-axis.
The geodesic motion was considered mainly in the
$(x^1,~x^2)$ plane only, $i.e.$ $x^\mu=0$, for $\mu>2$. 
Special attention has been paid to the case $D=5$.
For $D=4$, our results are in very good agreement 
with those in \cite{Dettmann:1994dj,Dettmann:1995ex}.

The trajectories are evolved numerically until an outcome is reached.
For a two black hole system, we distinguish three possible outcomes:
a test particle may reach the first  black hole,  the second black hole 
or may orbit indefinitely.

This last possibility has two different subclasses: 
there are particles which escape to infinity 
and particles that get confined  
and follow indefinitely an orbit in a limited region
\newpage
\setlength{\unitlength}{1cm}

\begin{picture}(18,7.5)
\centering
\put(2,0.0){\epsfig{file=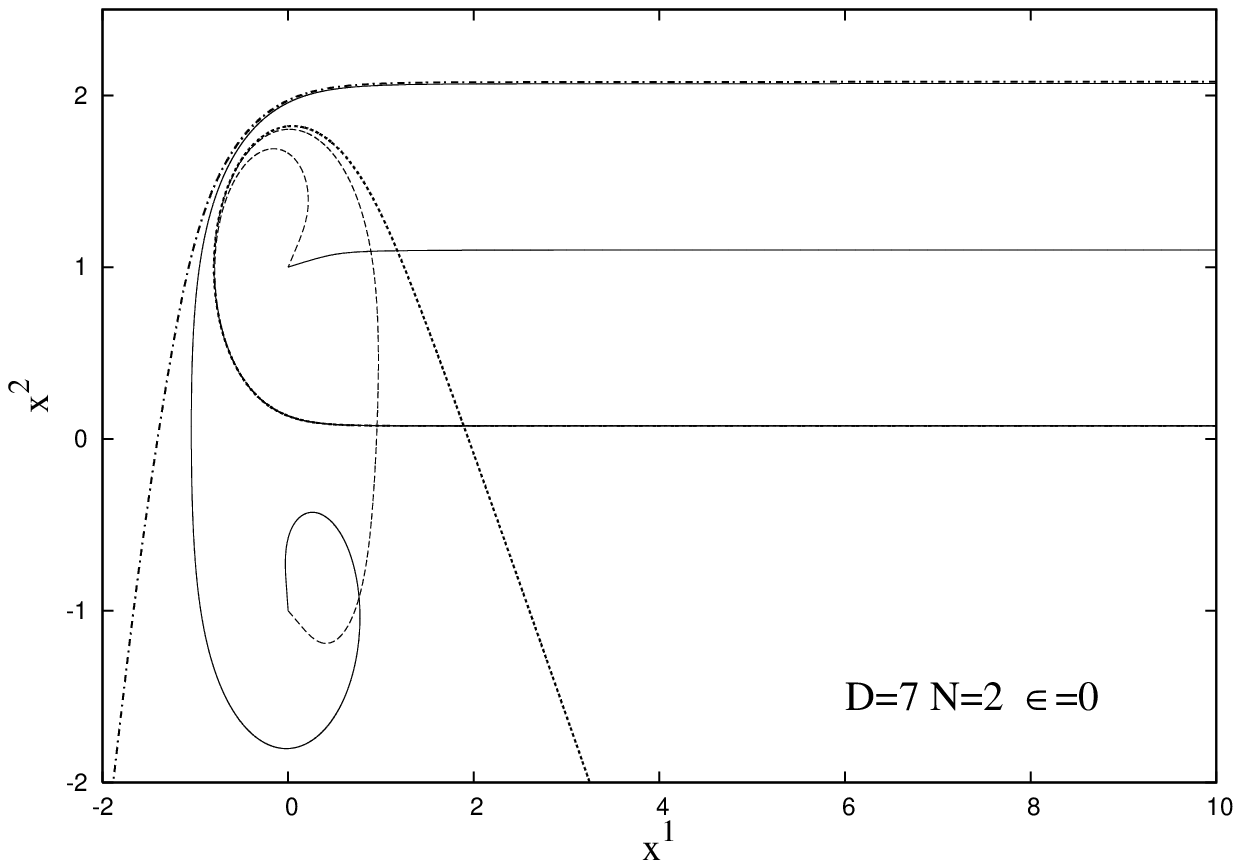,width=12cm}}
\end{picture}
\\
\\
{\small {\bf Figure 1.}
Typical null trajectories are shown  for a $D=7$, $N=2$ MP black hole.
The black holes are located on the $x^2$-axis, at $+1$ and $-1$ respectively.
The test particles start at $x^{1(0)}=10$
with $u_i^{(0)}=-10\delta_i^1$.
 }
\\
\\
 of spacetime.
Typical null trajectories illustrating these outcomes
are shown in 
Figure 1 for a $D=7$ $N=2$ system,
the orbits of massive particles presenting a rather similar shape. 

A ${\bf u}^{(0)}=0$ slice of phase space for 
$D=5,~N=2$ timelike test particles is presented in 
Figure 2.
The two black holes there have equal masses $M_i=0.01$ and 
are placed at $(0,-1)$ and $(0,1)$ respectively.
The initial location of the test particle is painted black if it ends 
 at $(0,-1)$ and white if it hits   the second black hole at $(0, 1)$.

One can see that 
these pictures look similar to those exhibited in the literature for 
the $D=4$ case
\cite{Dettmann:1994dj,Dettmann:1995ex}.
In particular,
the boundary between the basins of attraction clearly has a complicated structure,
which may be quantified using the concept of fractal dimension
(see \cite{ Dettmann:1995ex,Cornish:1996de} for other possible approaches).

To estimate the fractal dimension of
the basin boundary we use the definition
\begin{equation}
d_{B}=\lim_{ \varepsilon \rightarrow 0}\frac{\ln N(\varepsilon)}{\ln (1/\varepsilon)}~,
\end{equation}
where $N(\varepsilon)$ is the minimal number of boxes 
of size $\varepsilon$ needed to cover the fractal set.
In practice one estimates $d_{B}$ by plotting $\ln N(\varepsilon)$ vs. $\ln  \varepsilon$.
The value of $d_{B}$ evaluated for the data in Figure 2 on a grid  with $2500^2$ points
is $d_B=1.34\pm 0.07$.
This noninteger value  shows in a coordinate invariant way that the
basin boundary is indeed a fractal.

A qualitatively similar picture  was found for all configurations we have considered, 
$i.e.$ we did not notice the existence of a critical spacetime dimension,
beyond which, for example, the structure cease to be fractal in nature.
 However, our findings indicate that, for a given 
$(D,~N)$, the fractal dimension
 is configuration dependent, even for ${\bf u}^{(0)}=0$.
That is, changing  the mass parameter or the location of the black holes leads to
different values of 
\newpage
\setlength{\unitlength}{1cm}
 
\begin{picture}(18,7)
\put(-1.2,0){\epsfig{file=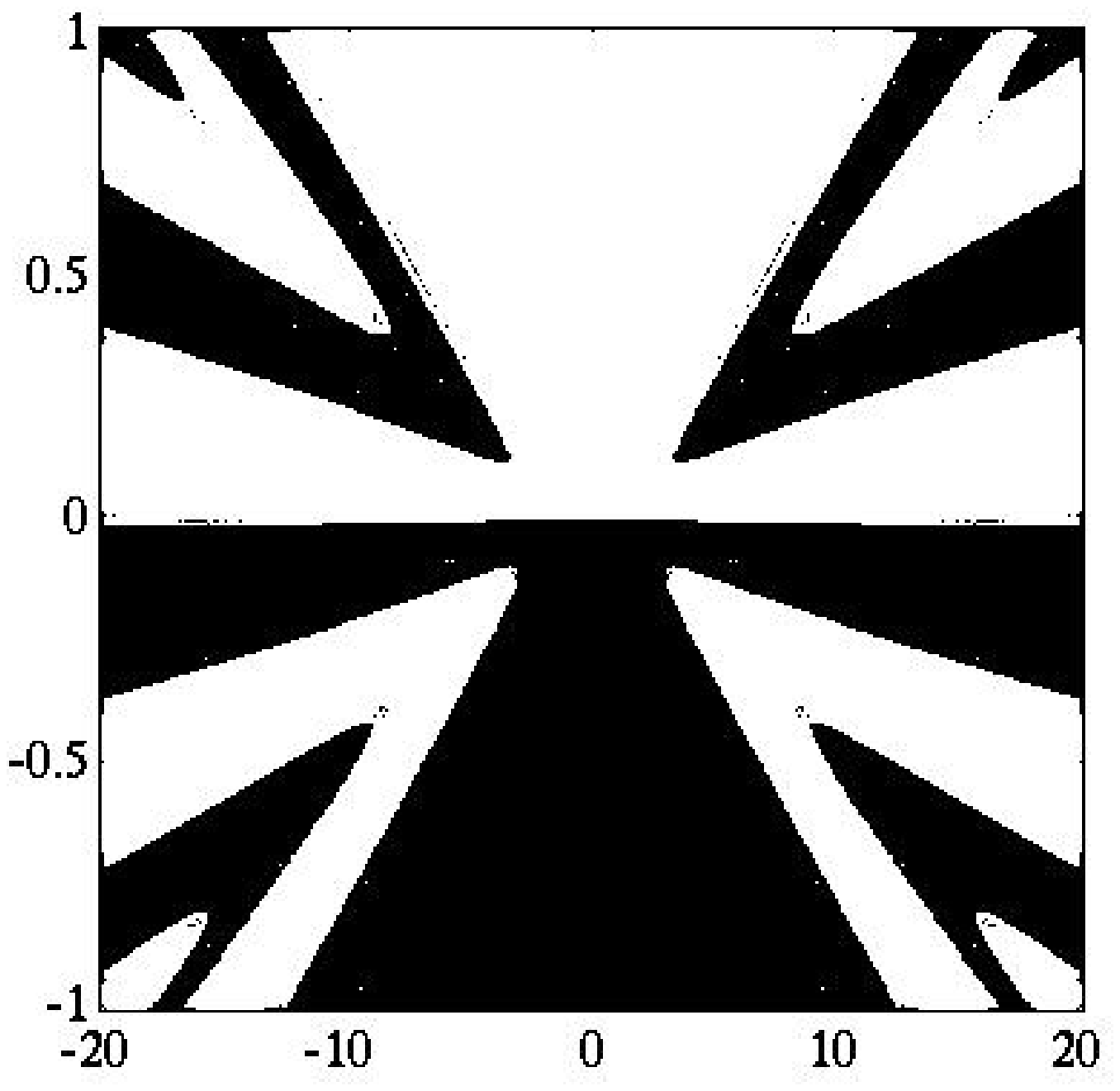,width=8.cm}}
\put( 7.3,0){\epsfig{file=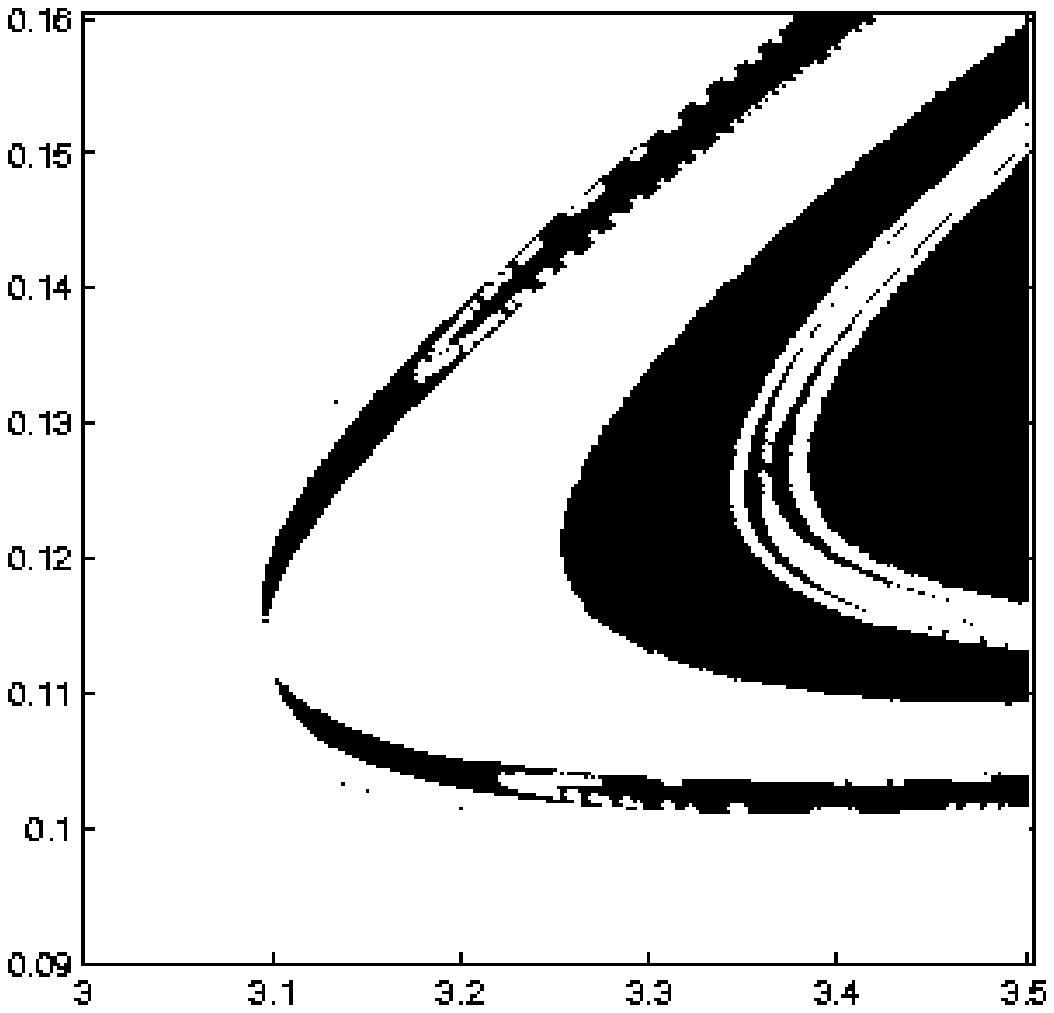,width=8.1cm}} 
\end{picture}
\\
\\
{\small {\bf Figure 2.}
A ${\bf u}=0$ section of phase space for $D=5,~N=2$  timelike trajectories.
The points here are colored according to their final outcome.
Here and in Figure 5, the second plots show the fractal structure 
for a small region of the first
pictures. Note also that in both figures 
the horizontal and the vertical axes correspond to the $x^1$ and $x^2$ axes, respectively.}
\\
\\
 $d_{B}$. Also, increasing 
the spacetime dimension makes  it
more difficult to analyse the fractal structure of the boundaries, 
the size of the relevant 
phase space regions generally decreasing  with $D$.

{\bf Holographic screens--}
An interesting physical question is how these  features will reflect
on the properties of a screen map of the multi-black hole system.
The notion of screen mapping was introduced by Susskind in Ref. \cite{Susskind:1994vu}
for $D=4$ asymptotically flat spacetimes,
in an attempt to implement `t Hooft's holographic
hypothesis \cite{'tHooft:1993gx}. 
In this approach, all points of space are mapped by light rays that impinge
perpendicularly onto a flat two-dimensional  screen  in a
distant asymptotically flat region.  
According to the 
holographic hypothesis, the black hole has the maximal bit density
of one per unit area. 
Thus the
horizons of any black holes are necessarily mapped
onto sets of larger area  on the screen. 
Corley and Jacobson refined the idea of Susskind and clarified the 
global properties of such maps \cite{Corley:1996qh}.
This concept  has proven seminal for 
further developments in the area of the
holographic
principle \cite{Bigatti:1999dp}, \cite{Bousso:1999cb}, \cite{Boyda:2002ba}.

The properties of the screen map of a single black hole
have been discussed in \cite{Corley:1996qh}.
Ref. \cite{Jackson:2001tg} dealt with two 
Schwarzschild black holes 
in the limit of large separation, in which case some of 
the interesting features 
of the geodesic
motion present in the MP case are absent (also, this configuration
 presents singularities on the axis joining the black holes \cite{ik,Emparan:2001bb}).

 Following \cite{Susskind:1994vu}, we have considered future directed light 
rays orthogonally to a screen placed
at some distance $L$ from the origin (with $L$ between $10$ and $10^3$).
(Here we will restrict to 
the case $D=4,~N=2$ and note $x^1=x,~x^2=y,~x^3=z$. 
The black holes are 
located  at $\pm 1$ on the $y-$axis and 
have equal masses $M_i=1/3$, the configuration
presenting 
\newpage
\setlength{\unitlength}{1cm}

\setlength{\unitlength}{1cm}
\begin{picture}(18,7)
\put(-1.4,-0.45){\epsfig{file=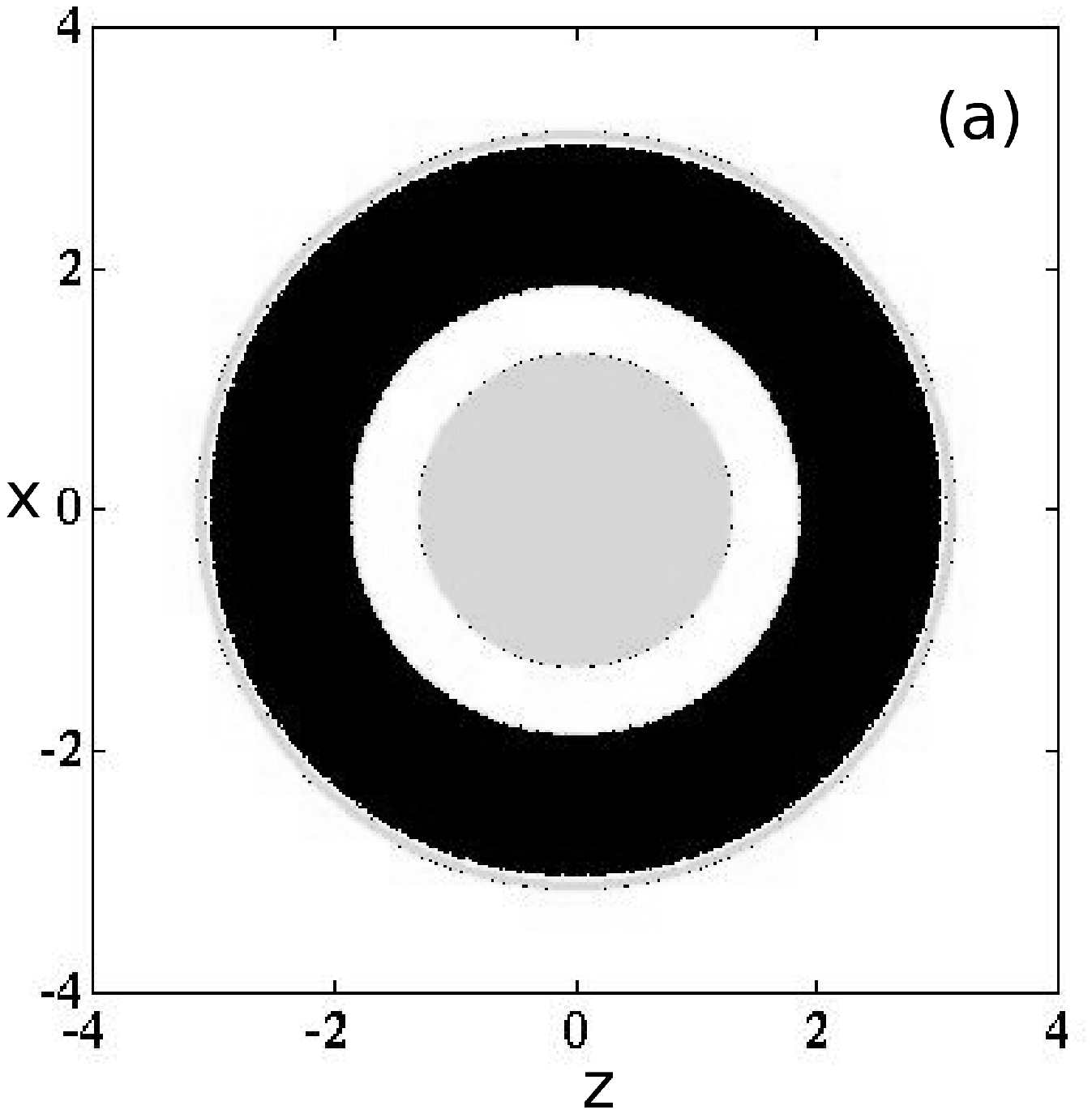,width=8.7cm}}
\put( 7.1,-0.38) {\epsfig{file=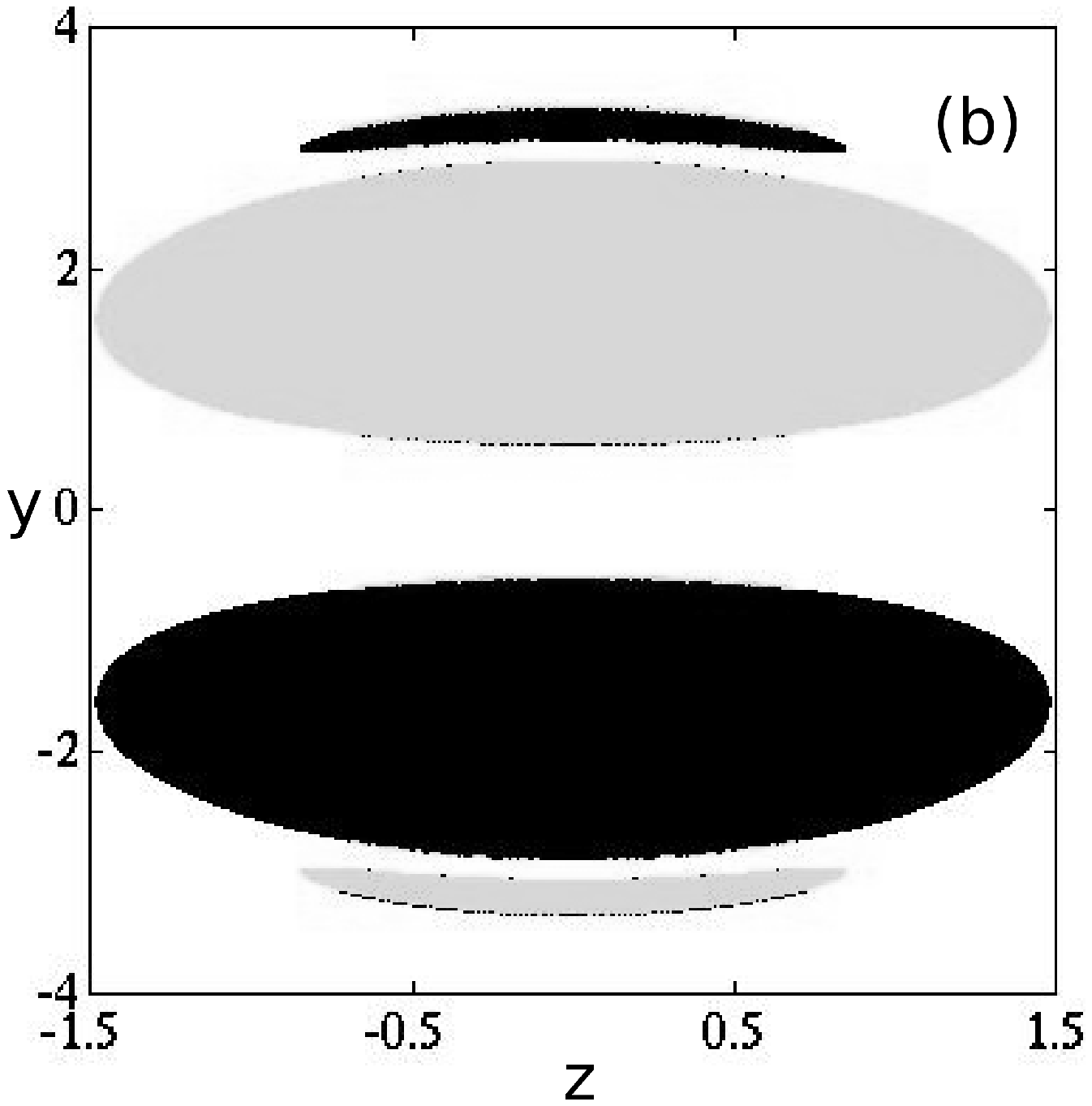,width=9.cm}}  
\end{picture}
\\
\\
{\small {\bf Figure 3.}
Screen images of a $D=4,~N=2$ system with $M_i=1/3$. 
The first black hole is located at $(0,-1,0)$   and the
second one at $(0,1,0)$. 
 The screen is placed at $L=100$, the light rays
 being perpendicular to the screen with $|\bf{u}^{(0)}|$=10.
 In these pictures, black stays for a particle being captured by the 
first black hole, gray for the second one and white for 
particles which orbit  indefinitely.
 }
\\
\\
an azimuthal symmetry).
The black holes image on the screen depends on the angular  
location of the screen with respect to the axis joining the black holes.
However, we have 
found that, for any screen location, the black holes have 
distinct images.
For a screen  
which lies perpendicular to the $y$-axis
(Figure 3a),
the image of the second black hole forms a ring around the first.
Apart from these primary images, there is also a sequence of 
concentric rings with decreasing area.
The primary black hole images on a screen perpendicular to the the $x$-axis 
are deformed disks (Figure 3b).
Secondary images are found at the $y-$extremities of these disks.
 The primary map of an individual black hole  
has an area which is greater than the horizon 
area, which gives an entropy density on the screen less than 1/4G, as expected
\cite{Susskind:1994vu}. 

In both cases above, zooming the boundary between the region with different outcomes on the screen map
reveals a complicated  structure, 
and  we notice the existence of a (presumably-) infinite sequence
of images with decreasing area.

To understand some of these features, one should remark that the generic
picture presented in Figure 2  remains the same for 
very large values of the radial coordinate. In particular,
it appears that the fractal boundary there (and the corresponding one for 
null motion) continues to spacelike infinity.
Also, the data on a planar screen represents a slice of the bulk phase space.

A crucial point here is the observation that when cutting 
a fractal structure in  a plane 
with a 
one dimensional line, self-similarity implies that
the fractal dust on that line has a
 fractal dimension $d_B-1$ \cite{Mandelbrot}.
For example, as seen in Figure 4, a fractal structure is induced 
on a circle of radius $r_0$ in the $(x,~y)$-plane (with $r_0$ much greater than the
 distance between 

\newpage
\setlength{\unitlength}{1cm}

\begin{picture}(18,7.2)
\centering
\put(2,0.0){\epsfig{file=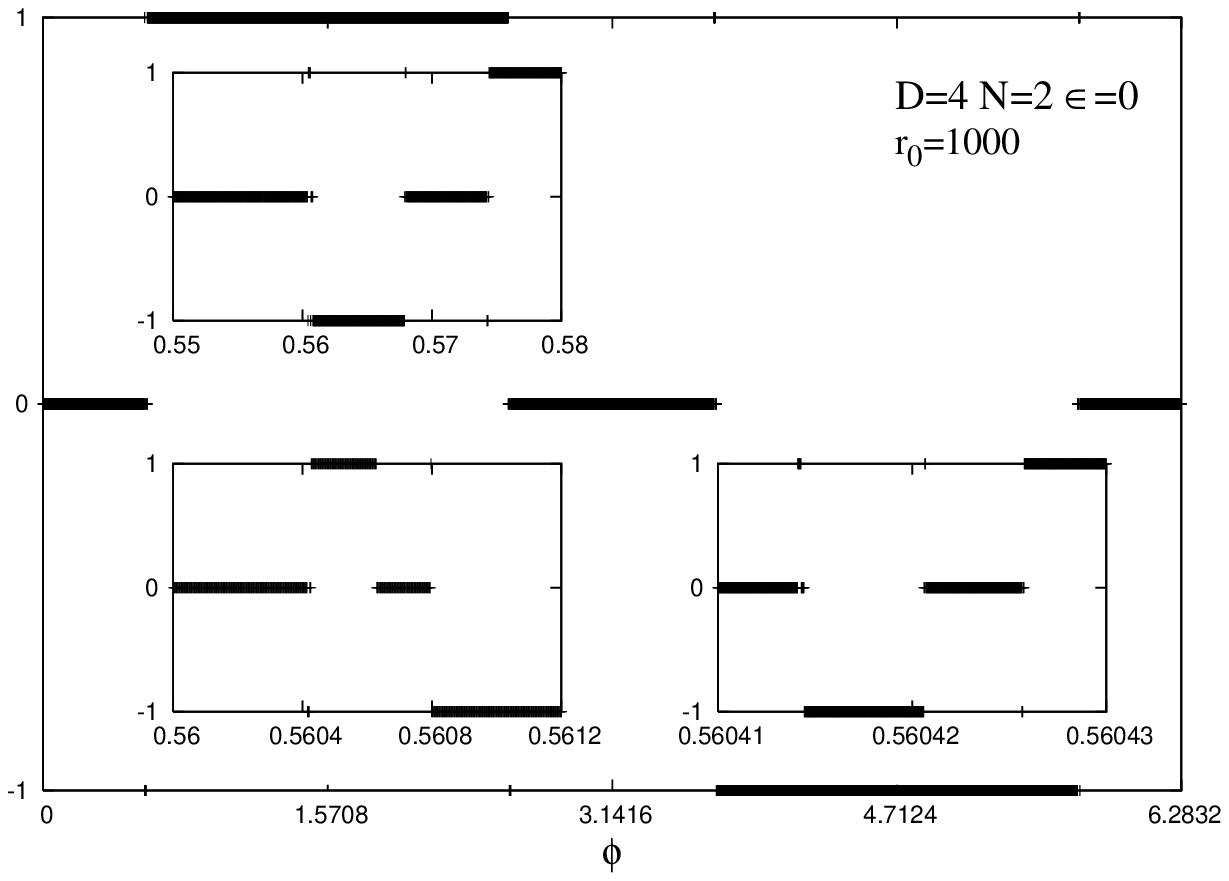,width=12cm}}
\end{picture}
\\
\\
{\small {\bf Figure 4.}
The fractal structure induced on a circle of radius $r_0$. 
 $\phi$ is the coordinate on the circle in the $(x,y)$-plane, with $\phi=0$
corresponding to $x=0$.
The vertical axis shows the outcome of the test particles, 
$-1$ for the black hole at $(0,-1)$,  $-1$ for the black hole at $(0,1)$ and
$0$ for those particles which orbit indefinitely.
In the close up views, we show transition regions between various outcomes, each successive
view
resolving a portion of the previous close up.
 }
\\
\\
 the black holes).
The photons start orthogonally to the circle,
 $|{\bf u}^{(0)}|=10$, directed inwards.
Zooming a transition region in that plot reveals a self similar 
structure;
the fractal 
structure was tracked over 12 decades in magnitude
before saturating 
the machine precision. 

Since a planar screen can be considered  as
a portion of a large sphere, 
this implies the 
existence of a fractal 
structure on the planar screen too. 
Thus, an angular slice along some direction on the screen plane reveals the existence of
a self similar structure
with a fractal dimension $d_B-1$, as illustrated in Figure 4. 

It is tempting to interpret this feature as a consequence of the holographic principle;
studying the black hole images on some screen provides information on 
the phase structure of the bulk spacetime.

Although we restricted ourselves here to the $D=4$ case, it is likely that
this is a generic property of the geodesic motion in a MP spacetime for any spacetime dimension.
Similar results have been found for $D=5,~N=2$ configurations.

{\bf Further remarks--}
The investigation of the geodesics in a higher dimensional
MP multi-black hole spacetime revealed that the features 
of the four dimensional case appear to be generic,
 the boundary separating in phase space the different
 asymptotic behaviours being fractal.
As usual, the presence of a fractal boundary indicates that there are nonsmooth
structures in phase space, which implies that the system is chaotic. 

The analysis in this paper can be straightforwardly extended to 
 test particles with an $U(1)$ electric charge $q$ and mass $m$.
We expect that similar to the situation in $D=4$,
the structure of the phase space will depend 
only slightly on the particle charge, except for the 
extremal
case $q=m$.

 \newpage
\setlength{\unitlength}{1cm}

\begin{picture}( 18,7)
\put(-1.2,0.2){\epsfig{file=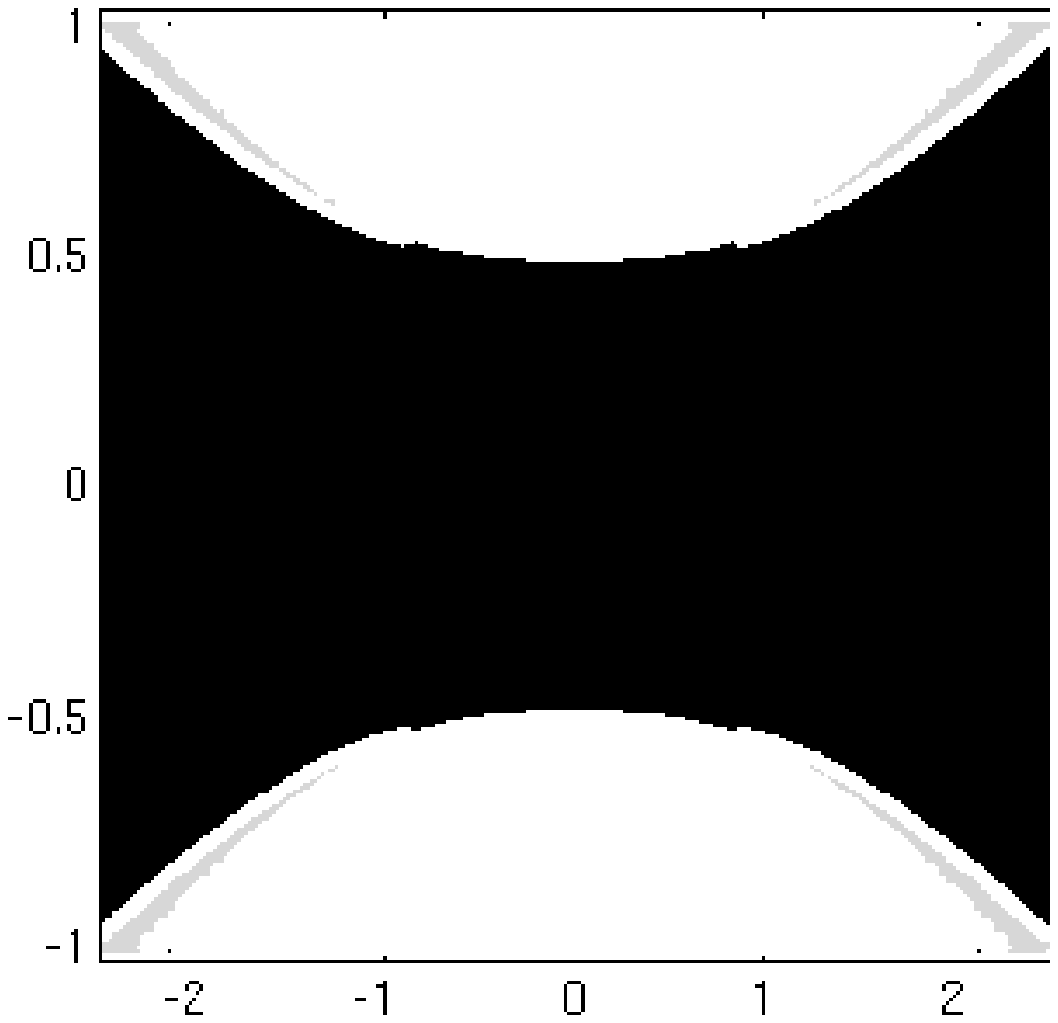,width=7.8 cm}}
\put(7.3,0){\epsfig{file=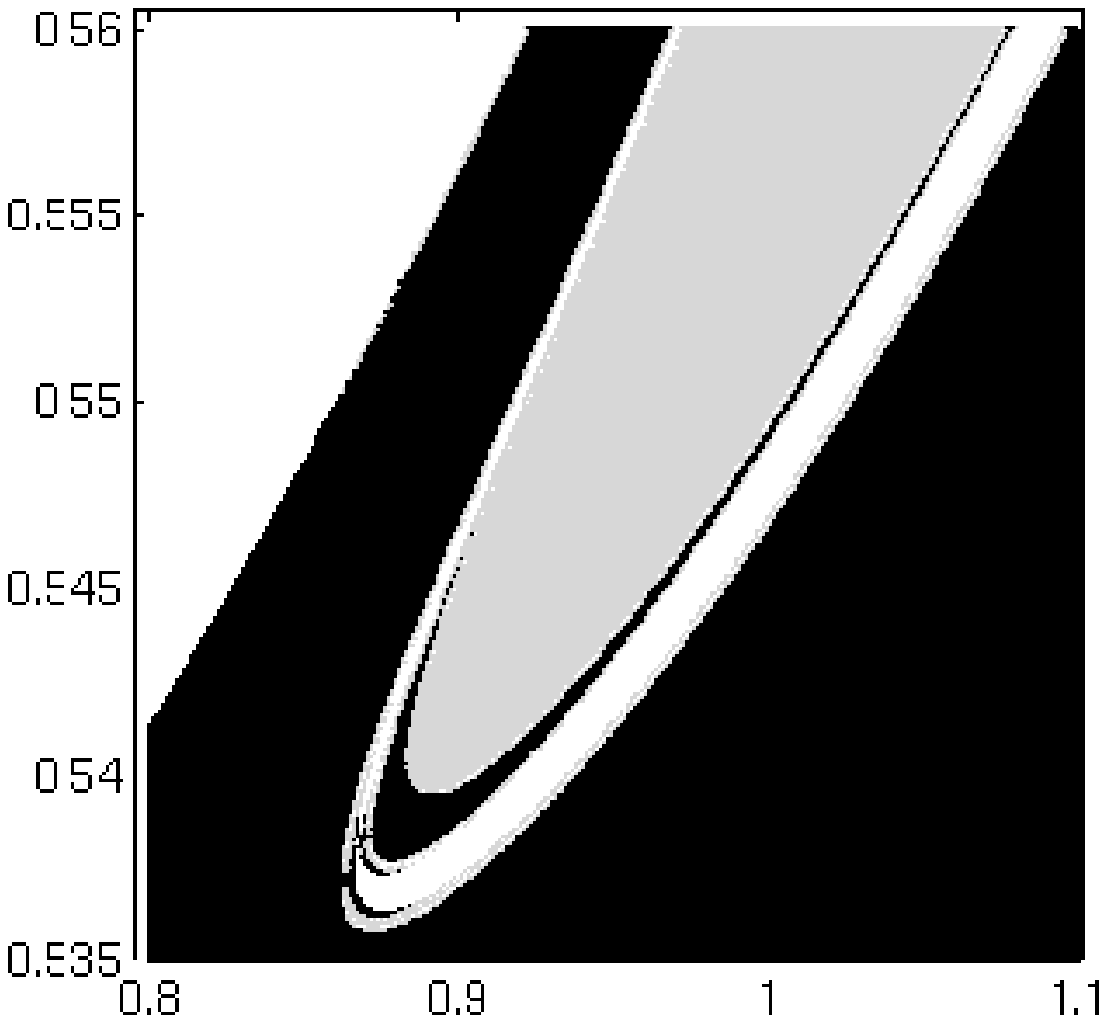,width=8.25cm}}  
\end{picture}
\\
\\
{\small {\bf Figure 5.}
A ${\bf u}=0$ section of phase space for $D=5, N=3$  timelike trajectories.
The black holes are located on the $x^2$-axis, at $\pm 1$ and $0$ respectively,
and have $M_i=0.01$. 
 The basin of attraction of the black hole at $x^2=-1$ is marked in black,  
 the  basin of attraction of the black hole at $x^2=1$ is marked in white, while those 
 trajectories ending on the  black hole at $x^2=0$ are marked in gray.
 }
\\
\\
Concerning the configurations
with several black holes,
we have found that the $N=2$ 
pattern repeats 
up to the $N=4$ case, the largest number of black holes we considered.

For the  case  with several black holes placed on the $x^2$-axis,
the basin of attraction still has a fractal structure.
To illustrate these aspects,  we give in Figure 5 the
results for timelike motion with ${\bf u}^{(0)}=0$ in a $D=5,~N=3$ MP spacetime. 
The points in that grid are again colored according to their
final outcome.

It is interesting to note that
if we place an infinite number of black holes along a line 
with each a unit distance apart, the resulting
spacetime has a periodicity one along the line.
This procedure is equivalent to compactifying one dimension
on a torus with period one. The resulting solution describes
a Kaluza-Klein extremal black hole, presenting  nontrivial dependence on the 
extra-dimension \cite{Myers:1986rx}.
Our results suggest that the geodesic motion 
in this background would also present chaotic features.
 However, a study of this case requires a separate discussion and 
is beyond the purposes of this work.
 
Although further work is clearly necessary,
we have pointed out a curious relation between the bulk phase structure and the induced
map on a holographic screen placed in the asymptotic
region. An observer on the screen would notice a fractal structure of the
corresponding phase space, whose fractal dimension is the bulk fractal dimension
minus one.

We would like to point out that this appears to be a generic feature 
of gravity solutions presenting chaotic features, the MP spacetime considered here being only
one particular case.
A similar approach can be applied to other spacetimes presenting chaotic motion.
It would be 
interesting to further explore the connection between the holographic 
principle and a fractal structure in the bulk.
 \\
{\bf  Acknowledgements} \\
  The authors thank Professor D. M. Heffernan  for useful discussions.
The work of ER is carried out
in the framework of Enterprise--Ireland Basic Science Research Project
SC/2003/390 of Enterprise-Ireland. 
 \begin{small}
 
 \end{small}

\end{document}